\documentclass[%
 reprint,
superscriptaddress,
frontmatterverbose, 
 amsmath,amssymb,
 aps,
prb,
]{revtex4-2}

\usepackage{graphicx}
\usepackage{dcolumn}
\usepackage{bm}
\usepackage[usenames,dvipsnames]{xcolor}
\usepackage{multirow}
\usepackage{makecell}
\usepackage{amsmath}
\usepackage{placeins}
\usepackage{soul}
\usepackage{ulem}
\usepackage{hyperref}
\usepackage[mathlines]{lineno}

\hypersetup{hidelinks}

\usepackage{xurl}

\begin{document}

\preprint{APS/123-QED}

\title{Spin-lattice model simulations of tetragonal FePt system } 

\author{Jakub Šebesta}
\email{jakub.sebesta@vsb.cz}
\affiliation{IT4Innovations, VSB – Technical University of Ostrava, 17. listopadu 2172/15, 708 00 Ostrava, Czech Republic}%

\author{Dominik Legut}
\affiliation{Charles University, Faculty of Mathematics and Physics
Ke Karlovu 3, 121 16 Praha 2, Czech Republic}%
\affiliation{IT4Innovations, VSB – Technical University of Ostrava, 17. listopadu 2172/15, 708 00 Ostrava, Czech Republic}%

\date{\today}

\begin{abstract}
Magnetic materials play a key role in the contemporary industry, providing unique features with a wide application potential. To study physical phenomena and design new materials, it is important to possess an appropriate tool, a model allowing  simulation of desired behavior.  Spin-lattice model simulations can be used to investigate spacious systems containing thousands of atoms, while complex phenomena arising from the interplay of lattice and spin dynamics can be modeled. One of the important phenomena that can be modeled in the spin lattice simulation is magnetoelastic behavior, offering direct conversion between the mechanical and magnetic energy. However, so far, only models for systems with cubic symmetry have been introduced. Therefore, here, a spin-lattice model for a system with tetragonal symmetry is proposed, where its strength is manifested by simulation of magnetoelastic properties of a characteristic representative L1$_{0}$ FePt system.
\end{abstract}

\maketitle


\section{Introduction}

Magnetic materials are widely used in daily life, however simulations of their behavior are much more complicated than for the non-magnetic systems.  For certain physical phenomena, the magnetic and elastic properties cannot be treated separately as the interplay between magnetic and mechanical degrees of freedom plays a vital role. Such coupling is called magnetoelasticity merging magnetic and elastic properties of the system~\cite{chikazumi2009physics,MagnetoelasticWaves_book}. It is manifested, e.g., in axial or volume magnetostriction and it allows a conversion between magnetic and mechanical energy which is employed in various sensors and actuators e.g., torque sensors, fuel injectors, micro-pump, weak-field detectors etc.~\cite{nishibe_r98_torquq,garshelis_r99,Bestok_micropump,PETRIDIS2006131,Spetzler_scirep_r21,Calkins_jimss_r07,Bienkovsky_SensAct_r04}. 

To simulate the magnetoelastic coupling, distinct methods can be used depending on the system size. For small systems, a method combining atomistic spin dynamics and ab-initio molecular dynamics (ASD-AIMD)~\cite{Stockem_prl_r18_asdaimd} might be used. However, such an approach became computationally demanding for large system based on a spacious suppercell, where the spin-lattice (SL) model~\cite{Cooke_cpc_r22_slmodel,Tranchida_jpc_18_spinpackage} based simulations take place. The SL model considers at once the spin and lattice degrees of freedom. The SL dynamics is governed by a Hamiltonian considering together the interatomic potential, magnetic interactions, such as the Heisenberg or dipole-dipole interactions, magnetocrystalline anisotropy, etc. The SL model offers a modeling system behavior in time and at finite temperature, which allows one to study e.g., magnon-phonon interactions, ultra-fast demagnetization processes, and energy flows in the lattice and spin subsystems.

The SL model has been successfully employed for a description of magnetoelastic behavior of ferromagnetic (FM) cubic systems~\cite{Nieves_prb_21_sl_model,Korniienko_PRR_r24,Korniilenko_ResPhys_r25}. However, their magnetoelastic behavior is weak~\cite{MAELAS_2_r22}. Much stronger magnetoelastic response can be observed in tetragonal systems due to the symmetry, as for L1$_{0}$ FePt system~\cite{Nieves_sss_r25_seconorder} possessing substantial magnetocrystalline anisotropy~\cite{Nieves_sss_r25_seconorder,Ravindran_prb_r01}.  
Therefore, in this paper, the spin-lattice model description for the tetragonal symmetry is derived. The SL model is applied to simulate the magnetoelastic behavior of the tetragonal FePt system.

\section{Theory and Methodology}

For magnetic systems, besides the elastic energy $E_{el}$, magnetoelastic energy $E_{me}$ describing the interplay between the magnetic and mechanical degrees of freedom has to be taken into account. Regarding the tetragonal (I) symmetry, the magnetoelastic energy in Cartesian reads~\cite{MAELAS_1_r21}
\begin{align}
\label{Eq.E_mag-elast}
   \frac{1}{V_{0}} E_{me} &=b_{11}(\varepsilon_{\mathrm{xx}} + \varepsilon_{\mathrm{yy}})  + b_{12}\varepsilon_{\mathrm{zz}}  \\ & + b_{21} (\alpha_{z}^{2} - \frac{1}{3})(\varepsilon_{\mathrm{xx}} + \varepsilon_{\mathrm{yy}} ) +  b_{22} (\alpha_{z}^{2} - \frac{1}{3})\varepsilon_{\mathrm{zz}} \nonumber \\ & +  \frac{1}{2} b_{3} (\alpha_{x}^{2} - \alpha_{y}^{2}) (\varepsilon_{\mathrm{xx}} - \varepsilon_{\mathrm{yy}}) + 2 b_{3}^{\prime} \alpha_{x} \alpha_{y} \varepsilon_{\mathrm{xy}}  \nonumber  \\ &+ 2 b_{4} ( \alpha_{x} \alpha_{z} \varepsilon_{\mathrm{xz}} + \alpha_{y} \alpha_{z} \varepsilon_{\mathrm{yz}}) \nonumber \, ,
\end{align}
where $\varepsilon_{ij}$ denotes components of strain tensor $\bm{\varepsilon}$, $\alpha_{i}$ stands for Cartesian components of the magnetization directions ($\vert \bm{\alpha}\vert=1$) and $b_{i}$ represents magnetetoelastic constants, where the $b_{11}$ and $b_{12}$ constants are related to the isotropic volume, the rest of them describes the anisotropic behavior.

Considering the elastic energy $E_{el}$ for the tetragonal (I) symmetry~\cite{AELAS_r17}
\begin{align}
    &\frac{1}{V_{0}}( E_{\mathrm{el}}-E_{0}) =    \\
    &=\frac{1}{2}C_{11}(\tilde{\varepsilon}_{\mathrm{1}}^{2} + \tilde{\varepsilon}_{\mathrm{2}}^{2}) + C_{12}\tilde{\varepsilon}_{\mathrm{1}}\tilde{\varepsilon}_{\mathrm{2}} + C_{13}(\tilde{\varepsilon}_{\mathrm{1}} + \tilde{\varepsilon}_{\mathrm{2}}) \tilde{\varepsilon}_{\mathrm{3}} \nonumber \\
    &+ \frac{1}{2}C_{33} \tilde{\varepsilon}_{\mathrm{3}}^{2} +\frac{1}{2} C_{44}\tilde{\varepsilon}_{\mathrm{4}}^{2} + \tilde{\varepsilon}_{\mathrm{5}}^{2} + \frac{1}{2}C_{66} \tilde{\varepsilon}_{\mathrm{6}} \nonumber \\
    &=\frac{1}{2}c_{xxxx}(\varepsilon_{\mathrm{xx}}^{2} + \varepsilon_{\mathrm{yy}}^{2}) + c_{xxyy}(\varepsilon_{\mathrm{xx}}  \varepsilon_{\mathrm{yy}}) + c_{xxzz}(\varepsilon_{\mathrm{xx}} + \varepsilon_{\mathrm{yy}}) \varepsilon_{\mathrm{zz}} \nonumber \\
    &+ \frac{1}{2}c_{zzzz} \varepsilon_{\mathrm{zz}}^{2} +2 c_{yzyz}(\varepsilon_{\mathrm{yz}}^{2} + \varepsilon_{\mathrm{zx}}^{2}) + 2c_{xyxy} \varepsilon_{\mathrm{xy}} \nonumber \, ,
    \label{Eq.Eel_tetragonal}
\end{align}
and the equilibrium strain is defined as follows
\begin{equation}
    \frac{\partial  E_{el} + E_{me}}{\partial \varepsilon^{{eq}}_{ij}} = 0  \, ,
\end{equation}
the magnetostriction described by the relative length change $\Delta l/l_{0}$ along the direction $\bm{\nu}$ for the magnetization direction $\bm{\alpha}$ reads
\begin{align}
    & \left.\frac{\Delta l }{l_{0}} \right\vert^{\bm{\alpha}}_{\bm{\nu}}  = \lambda^{\alpha 1,0} (\nu_{x}^{2} + \nu_{y}^{2}) 
    + \lambda^{\alpha 2,0} \nu_{z}^{2} \label{Eq.rel_l_change} \\  &  +  \lambda^{\alpha 1,2} (\alpha_{z}^2-\frac{1}{3}) (\nu_{x}^{2} + \nu_{y}^{2}) + \lambda^{\alpha 2,2} (\alpha_{z}^2-\frac{1}{3}) \nu_{z}^{2}
    \nonumber  \\
    &  +\frac{1}{2}  \lambda^{\gamma,2} (\alpha_{z}^2-\alpha_{y}^{2}) (\nu_{x}^{2}  - \nu_{y}^{2}) + 2\lambda^{\delta,2} \alpha_{x}\alpha_{y}\nu_{x}\nu_{y} \nonumber \\
    \nonumber & + 2\lambda^{\varepsilon,2} (\alpha_{x}\alpha_{z}\nu_{x}\nu_{z} + \alpha_{y}\alpha_{z}\nu_{y}\nu_{z}) \nonumber \, ,
\end{align}
where $\lambda^{i}$ denotes magnetostrictive coefficients.
Similarly to the magnetolelastic constants $b_{i}$, the isotropic $\lambda^{\alpha 1,0}$ and $\lambda^{\alpha 1,0}$ coefficients belong to the the volume magnetostriction~\cite{maelas3_nieves_r23}
\begin{equation}
        \lambda^{\alpha 1,0} = \frac{-b_{11}C_{33}+b_{12}C_{13}}{C_{33}(C_{11}+C_{12})-2C_{13}^2} \, ,
\end{equation}
\begin{equation}
        \lambda^{\alpha 2,0} = \frac{2b_{11}C_{13}-b_{12}(C_{11}+C_{12})}{C_{33}(C_{11}+C_{12})-2C_{13}^2} \, .
\end{equation}
The remaining ones are anisotropic coefficients related to the magnetization direction-dependent magnetostriction~\cite{MAELAS_1_r21} 
\begin{equation}
    \lambda^{\alpha 1,2} = \frac{-b_{21}C_{33}+b_{22}C_{13}}{C_{33}(C_{11}+C_{12})-2C_{13}^2}
\end{equation}
\begin{equation}
    \lambda^{\alpha 2,2} = \frac{2b_{21}C_{13}-b_{22}(C_{11}+C_{12})}{C_{33}(C_{11}+C_{12})-2C_{13}^2}
\end{equation}
\begin{equation}
    \lambda^{\gamma, 2} = \frac{-b_{3}}{C_{11}-C_{12}}
\end{equation}
\begin{equation}
    \lambda^{\delta, 2} = \frac{-b_{3}^{\prime}}{2C_{66}} 
\end{equation}
\begin{equation}
    \lambda^{\varepsilon, 2} = \frac{-b_{4}}{2C_{44}} \, .
\end{equation}

To model the magnetoelastic behavior of tetragonal systems, a spin-lattice Hamiltonian $H_{SL}$ with terms similar to the cubic model~\cite{Nieves_prb_21_sl_model,Tranchida_jpc_18_spinpackage} can be considered. It reads 
\begin{align}
    \mathcal{H}_{SL}(\mathbf{r},\mathbf{p},\bm{\alpha})= \sum_{i=1}^{N} \frac{\mathbf{p_{i}}}{2m_{i}} +  \sum_{i,j=1}^{N} \mathcal{V}_{ij}(r_{ij}) + \mathcal{H}_{M}(\mathbf{r},\bm{\alpha})\, ,
    \label{Eq.SL_Hamiltonian}
\end{align}
where $\mathbf{r}_{i}$, $\mathbf{p}_{i}$,
denotes the position, momentum 
of an atom $i$ with the mass $m_i$, $N$ is total number of atom,  $\mathcal{V}_{ij}(r_{ij})$ represents the interatomic potential dependent on the atom distance $r=|\mathbf{r_{i}}-\mathbf{r_{j}}|$ and magnetic interactions are described by the Hamiltonian $\mathcal{H}_{M}$. It includes the Zeeman term for the interaction with the external magnetic field,  magnetic exchange interaction of classical spins, pseudo-dipolar interaction, and tetragonal magnetocrystalline anisotropy (MCA) to include  effects arising from spin-orbit coupling~\cite{Tranchida_jpc_18_spinpackage}:
\begin{align}
\label{Eq.Ham_magnetic}
  \mathcal{H}_{M}&(\mathbf{r},\bm{\alpha})  = \\
  &=\mathcal{H}_{Z}(\bm{\alpha}) +  \mathcal{H}_{\mathrm{ex}}(\mathbf{r},\bm{\alpha}) +  \mathcal{H}_{\mathrm{di}}(\mathbf{r},\bm{\alpha}) +  \mathcal{H}_{\mathrm{MCA}}(\bm{\alpha}) \nonumber\\
  &-\mu_{0}  \sum_{i=1}^{N} \mu_{i} \mathbf{H}\cdot\bm{\alpha}_{i} -\frac{1}{2} \sum_{\substack{i,j=1 \\ i\neq j}}^{N} J_{ij}(r_{ij})\bm{\alpha}_{i}\cdot\bm{\alpha}_{j} \nonumber\\  
  &- \frac{1}{2} \sum_{i,j=1}^{N} l_{ij}(r_{ij})\Big[ (\mathbf{e}_{ij}\cdot\bm{\alpha}_{i})(\mathbf{e}_{ij}\cdot\bm{\alpha}_{j}) - \frac{(\bm{\alpha}_{i}\cdot\bm{\alpha}_{j})}{3} \Big]  \nonumber \\
  & + \Big[  K_{1}\mathrm{sin}^{2}(\theta) + K_{2}\mathrm{sin}^{4}(\theta) + K_{3}\mathrm{sin}^{4}(\theta)\mathrm{cos}(4\phi) \Big] \nonumber \, ; \\
    \mathbf{e}_{ij} & = \frac{\mathbf{r}_{ij}}{r_{ij}} \, ,
  \end{align}
where $\mu_0$ is the permeability of vacuum, $\mu_i$ denotes magnetic moment of the atom $i$, $H$ is an external magnetic field,   $J_{ij}(r_{ij})$ stands for magnetic exchange coupling, $l(r_{ij})$ is a parameter of the pseudo-dipolar interaction,
$K_{i}$ denotes magnetocrystalline anisotropy constants,
$\theta$ is the angle between the spin direction $\bm{\alpha}$ and tetragonal axes $c$ ($c\neq a=b$), and $\phi$ is an angle in the basal $ab$-plane between the spin direction $\bm{\alpha}$ and tetragonal axes $a$.

Performing the simulations within the LAMMPS package~\cite{LAMMPS,Tranchida_jpc_18_spinpackage,Cooke_cpc_r22_slmodel}, the radial dependence of the exchange interaction coupling $J_{ij}(r_{ij})$ and the pseudo-dipolar coupling  $l(r_{ij})$ can be  parametrized by a Bethe-Slater curve $\Phi(r_{ij})$~\cite{Tranchida_jpc_18_spinpackage}: 
\begin{align}
    \Phi(r_{ij}) = 4\alpha \Bigg( \frac{r_{ij}}{\delta} \Bigg)^2 \Bigg[1-\gamma\Big( \frac{r_{ij}}{\delta} \Big)^2 \Bigg] \mathrm{e}^{-  \Big(\frac{r_{ij}}{\delta}\Big)^2   } \Theta(R_{c} - r_{ij})\; , 
    \label{Eq.Bethe-Salpeter_curve}
\end{align}
where $\alpha$, $\gamma$, $\delta$ are parameters, and $\Theta(R_{c} - r_{ij})$ denotes the Heaviside step function with the radial cut-off $R_{c}$.

The anisotropic part of the magnetoelastic energy (Eq.~\ref{Eq.E_mag-elast}) can be included in the spin-lattice model Hamiltonian (Eq.~\ref{Eq.SL_Hamiltonian})  via the pseudo-dipolar term $\mathcal{H}_{di}$ (Eq.~\ref{Eq.Ham_magnetic})
\begin{align}
    &\mathcal{H}_{di}(\mathbf{r},\bm{\alpha})= \label{Eq.pseudodip_ham} \\
    &=- \frac{1}{2} \sum_{i,j=1}^{N} l_{ij}(r_{ij})\Big[ (\mathbf{e}_{ij}\cdot\bm{\alpha}_{i})(\mathbf{e}_{ij}\cdot\bm{\alpha}_{j}) - \frac{(\bm{\alpha}_{i}\cdot\bm{\alpha}_{j})}{3} \Big]\, , \nonumber
\end{align}
whereas the isotropic part can be modeled by the exchange interaction $\mathcal{H}_{ex}$.

For  FM ordered  spins ($\bm{\alpha}_{i}\cdot\bm{\alpha}_{j}=1$), the  pseudo-dipolar term $\mathcal{H}_{di}$ is  simplified as follows:
\begin{align}
    \mathcal{H}_{di}^{||}(\mathbf{r},\bm{\alpha})= - \frac{1}{2} \sum_{i,j=1}^{N} l_{ij}(r_{ij})\Big[ (\mathbf{e}_{ij}\cdot\bm{\alpha}_{i})^2 - \frac{1}{3} \Big]\, . \label{Eq.SLHamFM} 
\end{align}
Considering pseudo-dipolar interaction between various neighbors and its change under small deformations, the behavior of the tetragonal (I) magnetoelastic energy (Eq.~\ref{Eq.E_mag-elast}) can be reconstructed. To simplify the model and include only the magnetization direction dependent term in the magnetoelastic energy (Eq.~\ref{Eq.E_mag-elast}), the following form of the pseudo-dipolar interaction is considered:
\begin{align}
    &\tilde{\mathcal{H}}_{di}(\mathbf{r},\mathbf{s})=  \label{Eq.PsDipoleIModified} \\
    &= - \frac{1}{2} \sum_{i,j=1}^{N} l_{ij}(r_{ij})\Big[ (\mathbf{e}_{ij}\cdot\bm{\alpha}_{i})(\mathbf{e}_{ij}\cdot\bm{\alpha}_{j}) - \frac{(\bm{\alpha}_{i}\cdot\bm{\alpha}_{j})}{3} + \frac{1}{3
    } \Big]\, . 
    \nonumber
\end{align}

\begin{figure}
(a)\hspace{-2pt}
\includegraphics[width=0.9\columnwidth]{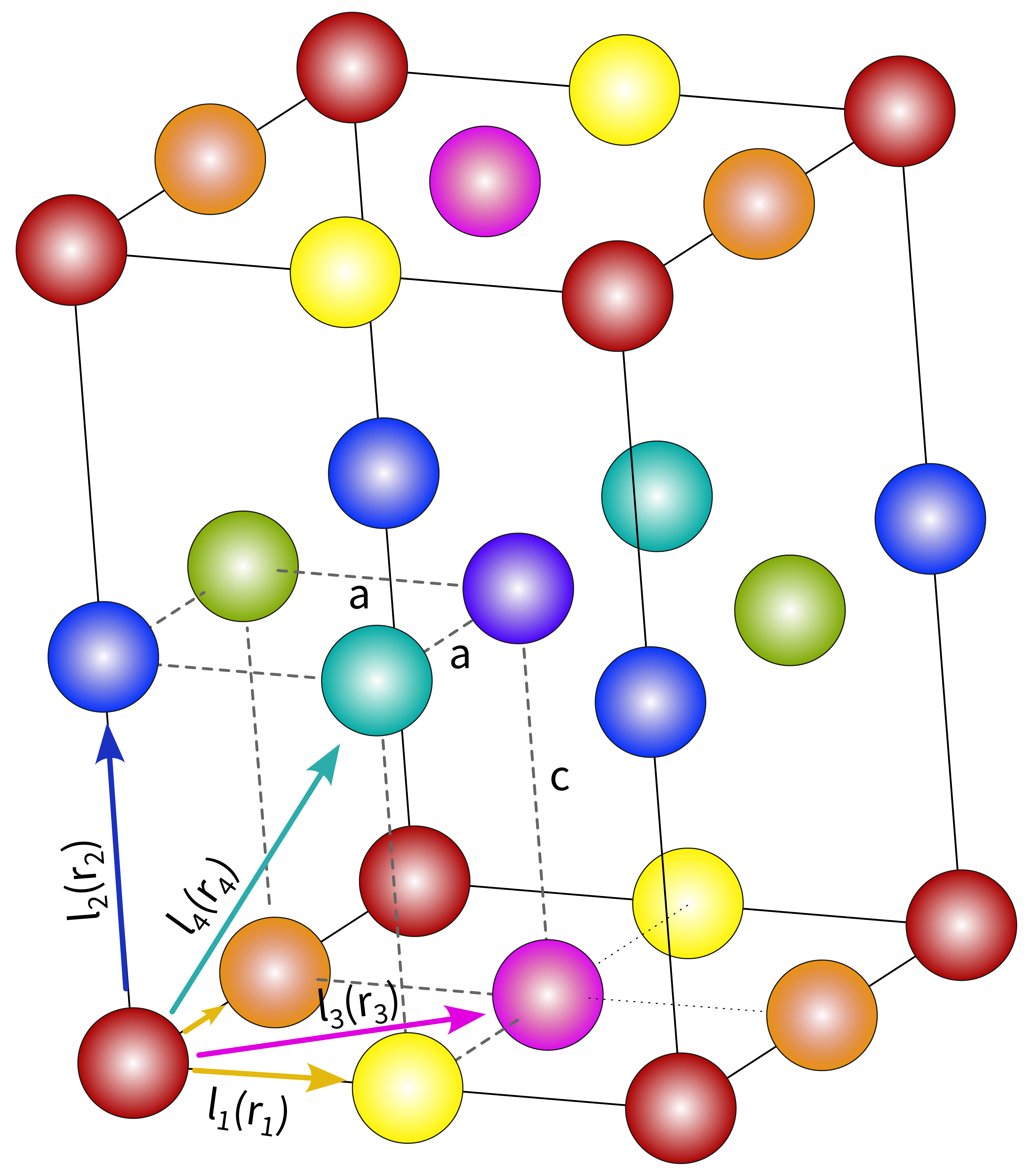}

\caption{\label{fig:Suppercel} Tetragonal supercell 2x2x2 with depicted pseudo-dipolar interactions $l_{i}(r_{i})$}
\label{Fig.:Tetr1_supcel222}
\end{figure}

Assuming  tetragonal (I) structure with a single atom in the primitive cell, FM ordered spins, and pseudo-dipolar interactions $\tilde{\mathcal{H}}_{di}$ (Eq. \ref{Eq.PsDipoleIModified})  up to the fourth-nearest neighboring atoms (Fig.~\ref{fig:Suppercel}), the change of  pseudo-dipolar Hamiltonian  $\Delta \tilde{\mathcal{H}}_{di}$ under small deformations, represented by the strain tensor $\bm{\varepsilon}$, can be approximate as follows:
\begin{widetext}
\begin{align}
    &-\Delta\tilde{\mathcal{H}}_{di}(\mathbf{r},\bm{\alpha},\bm{\varepsilon}) = \label{Eq.PsDipolarDiff} \\
    &=l_1(r_{01}) \, 2  \Big[ 2\alpha_{x}\alpha_{y}\varepsilon_{xy}  + \alpha_{x}\alpha_{z} {\varepsilon_{xz}}   +  \alpha_{y}\alpha_{z} {\varepsilon_{yz}}  \Big] +   r_{01}   \frac{\partial l_{1}}{\partial r} \Bigg\vert_{r=r_{01}}    \,  \Big[  \alpha_{x}^{2} \varepsilon_{xx} +  \alpha_{y}^{2} \varepsilon_{yy} \Big] \nonumber \\
    & +l_2(r_{02}) \,  2  \Big(\alpha_{x}\alpha_{z} \varepsilon_{xz} + \alpha_{y}\alpha_{z} \varepsilon_{yz} \Big) +  r_{02} \frac{\partial l_{2}}{\partial r} \Bigg\vert_{r=r_{02}}  \, \alpha_{z}^{2} \varepsilon_{zz}  \nonumber\\
    & +l_3(r_{03}) \, 2 \Big[    \alpha_{x}^{2} \varepsilon_{xx} +  \alpha_{y}^{2} \varepsilon_{yy} + 2 \alpha_{x} \alpha_{y} \varepsilon_{xy}  +  \alpha_{x} \alpha_{z} {\varepsilon_{xz}}  +  \alpha_{y} \alpha_{z} {\varepsilon_{yz}} \Big] \nonumber\\
    & + r_{03}   \frac{\partial l_{3}}{\partial r} \Bigg\vert_{r=r_{03}}  \,  \Big[   \alpha_{x}^{2} \varepsilon_{xx} +  \alpha_{x}^{2} \varepsilon_{yy} + \alpha_{y}^{2} \varepsilon_{xx} +  \alpha_{y}^{2} \varepsilon_{yy} + 4 \alpha_{x} \alpha_{y} \varepsilon_{xy}   \Big]\nonumber \\
 & +l_4(r_{04}) \,  \frac{4}{w^{2}} \Big[ \alpha_{x}^{2} \varepsilon_{xx} + \alpha_{y}^{2} \varepsilon_{yy} + 2\alpha_{x} \alpha_{y} \varepsilon_{xy}  +  \alpha_{x} \alpha_{z} \varepsilon_{xz} (2k^2 +1)    +  \alpha_{y} \alpha_{z} \varepsilon_{yz} ( 2k^{2} +1)  + 2 \alpha_{z}^{2} \varepsilon_{zz} k^{2}\Big]\nonumber \\
    &+r_{04}  \frac{\partial l_{4}}{\partial r} \Bigg\vert_{r=r_{04}} \frac{2}{w^{2}} \Big[ \alpha_{x}^{2} \varepsilon_{xx} + \alpha_{y}^{2} \varepsilon_{yy} + k^2 \Big(    4 \alpha_{x} \alpha_{z} \varepsilon_{xz}       + 4 \alpha_{y} \alpha_{z} \varepsilon_{yz}    \alpha_{z}^{2} \varepsilon_{xx} + \alpha_{z}^{2}  +  \alpha_{z}^{2} \varepsilon_{zz} (2k^2-1)  +   \varepsilon_{zz} \Big) \Big ] \, ;\nonumber \\
    &     k=\frac{c}{a}, w^2=(k^2+1)\, 
\end{align}

where $l_{i}(r_{i})$ denotes pseudo-dipolar interaction coupling between $i^\mathrm{th}$ nearest neighbors, $r_{0i}$ represents undeformed bond length $r_{i}$,  and $a$ resp. $c$ are tetragonal lattice parameters. 
\end{widetext}

Comparing the changes in the pseudo-dipolar interaction  under small deformations (Eq.~\ref{Eq.PsDipolarDiff}) to the form of the tetragonal magnetoelastic energy (Eq.~\ref{Eq.E_mag-elast}), the dipolar interaction coupling $l_{i}(r_{i})$ can be expressed in terms of the magnetoelastic constants $b_{i}$ in following way
\begin{align}
l_{1}(r_{01}) &=  \quad- \frac{{1}}{2}  b_{3}^{\prime} - {\frac{1}{8} b_{3}} &  \, , \label{Eq.psdip_l1}\\
 r_{01}\frac{\partial l_{1}}{\partial r} \Bigg\vert_{r=r_{01}}  & =  \quad  -{  b_{3}} + {\frac{ {4}}{ w^2k^2}b_{21}} &  \, , \label{Eq.psdip_dl1dr} 
\end{align}

\begin{align}
   l_{2}(r_{02}) & =  \quad-b_{4} +   \frac{{1}}{2}  b_{3}^{\prime}  + {\frac{ {8} }{ w^2}b_{21}}&\, , \label{Eq.psdip_l2}\\
 r_{02} \frac{\partial l_{2}}{\partial r} \Bigg\vert_{r=r_{02}}  & = \quad - b_{22} + {\frac{ {4} (2k^2-1) }{ w^2}b_{21}}& \, ,\label{Eq.psdip_dl2dr}
\end{align}

\begin{align}
l_{3}(r_{03})  &= \quad 0 & \, ,\label{Eq.psdip_l3}\\
   r_{03}\frac{\partial l_{3}}{\partial r} \Bigg\vert_{r=r_{03}}  &= \quad {\frac{1}{4} b_3} &\, ,\label{Eq.psdip_dl3dr}
\end{align}

\begin{align}
 l_{4}(r_{04}) &= \quad0  &  \, ,\label{Eq.psdip_l4}\\
 r_{04}\frac{\partial l_{4}}{\partial r} \Bigg\vert_{r=r_{04}} &= \quad -{\frac{ {2} w^2 }{k^2}b_{21}}& \,\label{Eq.psdip_dl4dr} ,
\end{align}
where mutual interrelations between the change of the bond length $r_{i}$ were taken into account. 
Assuming a primitive tetragonal 2x2x2 supercell (Fig.~\ref{Fig.:Tetr1_supcel222}), each type of pseudo-dipolar interaction $l_{i}$ can be attributed to specific pairs of interacting atoms. Then, regarding LAMMPS simulations, a pseudo-dipolar interaction $l_{i}(r_{i})$ can be parametrized by a Bethe-Slater curve $\Phi_{l,i}(r_{i})$  (Eq.~\ref{Eq.Bethe-Salpeter_curve}) where 
\begin{align}
    &\alpha_{l,i} = \frac{e}{8} \Bigg[ 2l_{i}(r_{0i}) - r_{0i}\frac{\partial l_{i}}{\partial r} \Bigg\vert_{r=r_{0i}}   \Bigg] \, , \\
    &\gamma_{l,i} =   \frac{r_{0i}\frac{\partial l_{i}}{\partial r} \Big\vert_{r=r_{0i}} }{r_{0i}\frac{\partial l_{i}}{\partial r} \Big\vert_{r=r_{0i}}  - 2l_{i}(r_{0i}) } \, ,\\
    &\delta_{l,i} = r_{0i}\, ,
\end{align}
and the radial cut-off $R_{c,l,i}$ is set to include only the nearest neighbors of the selected kind of interacting atoms. 

As the derived strain induced difference in   pseudo-dipolar interaction   (Eq.~\ref{Eq.PsDipolarDiff}) contains a magnetization direction independent term 
$r_{04}  \frac{\partial l_{4}}{\partial r} \Big\vert_{r=r_{04}} \frac{2 k^2}{w^{2}}     \varepsilon_{zz}$ , an extra interaction $\mathcal{H}_{\mathrm{off}}$ was introduced in the model to offset this term:
\begin{align}
   \mathcal{H}_{\mathrm{off}}(\mathbf{r})= - \frac{1}{2} \sum_{i,j=1}^{N} l_{ij}^{\mathrm{off}}(r_{ij})\, , \label{Eq.SLHamoffset} 
\end{align}
where
\begin{align}
  l^{\mathrm{off}}_{2}(r_{02})& =   \quad 0   \, , \label{Eq.psdip_off_l2}\\
r_{02}\frac{\partial l^{\mathrm{off}}_{2}}{\partial r} \Bigg\vert_{r=r_{02}} &=   \quad\frac{4}{w^2} b_{21}\, . \label{Eq.psdip_off_dl2dr}
\end{align}


For better feasibility of the model spin-lattice model, the  magnetocrystalline anisotropy  $\mathcal{H}_{\mathrm{MCA}}$  can be written as follows 
\begin{align}
    \label{Eq.MAE_Ham_sn}
    &\mathcal{H}_{\mathrm{MCA}}(\bm{\alpha}) = K_{1} \Big[1- (\bm{\alpha}\cdot\mathbf{n}_{c})^2\Big]  \\
    &+  \Big[1- (\bm{\alpha}\cdot\mathbf{n}_{c})^2\Big]^2 \Bigg[ K_{2} + K_{3} \Big[8(\bm{\alpha}\cdot\mathbf{n}_{a})^4-8(\bm{\alpha}\cdot\mathbf{n}_{a})^2+1\Big] \Bigg] \nonumber \, ; \\
    &\vert \mathbf{n}_{c} \vert  = \vert \mathbf{n}_{a} \vert = 1\nonumber
\end{align}
where $\mathbf{n}_{c}$ resp. $\mathbf{n}_{a}$ denote  directions of the tetragonal $c$- resp. $a$-axis.
Comparing the pseudo-dipolar interaction (Eq.~\ref{Eq.SLHamFM}) to the above expression of the MAE (Eq.~\ref{Eq.MAE_Ham_sn}), it is straightforward that the pseudo-dipolar interaction contributes to the MAE, namely the $K_{1}$ term. Therefore, the value of $K_{1}$ parameter in the SL model has to be corrected for the contribution of the pseudo-dipolar coupling. Then, the adjusted $\tilde{K}_{1}$ parameter reads
\begin{equation}
    \tilde{K}_{1} = {K}_{1} +2\Big(l_1(r_{01}) -l_2(r_{02}) +l_3(r_{03})\Big)  - \frac{4 l_4(r_{04})}{w^{2}} 
    \label{Eq.MAE_K1_off}
\end{equation}

\begin{table}[t]
\caption{\label{Tab:elast_prop}%
Elastic properties of L1$_{0}$ FePt obtained either by ab-initio VASP calculation or spin-lattice simulation in LAMMPS. Lattice parameters $a$ and $c$ relaxed by DFT calculations. Equilibrium volumes in the SL simulations are determined from the energy volume curves (Fig~\ref{fig:eos}).  }
\begin{ruledtabular}
\begin{tabular}{c|c|ccc}
 &VASP & \multicolumn{3}{c}{LAMMPS}  \\
 & \multirow{ 2}{*}{ PBE+SO} & \multirow{ 2}{*}{MEAM} & \multirow{ 2}{*}{RF-MEAM} & RF-MEAM  \\
  & &  &  & + SL model \\

\hline

$a$(\AA) & 2.723 \\
$a$(\AA) & 3.768\\
$V$(\AA$^3$) & 27.937 & 26.711 & 27.952 & 27.946 \\

\hline

C$_{11}$(GPa) &   374.6 &   353.5  & 365.7 &  365.6 \\
C$_{12}$(GPa) &   \phantom{0}81.4   &   185.2 &  \phantom{0}85.1 &  \phantom{0}85.1 \\
C$_{13}$(GPa) &   154.9   &   195.0 &  157.5 &  157.7  \\
C$_{33}$(GPa) &   299.4   &   245.3 &  299.4 &  300.8 \\
C$_{44}$(GPa) &   109.5   &   103.3 &  108.8 &   108.9 \\
C$_{66}$(GPa) &    \phantom{0}49.7   &   \phantom{0}35.7 &  \phantom{0}56.8 &   \phantom{0}56.9 \\
					
	\hline				
$B$(GPa) &203.4&233.6&203.4&203.7 \\

\end{tabular}
\end{ruledtabular}
\end{table}

\begin{table}[b]
\caption{\label{Tab:bi_Bethe-Slater_param}%
Bethe-Slater curve parameters for pseudo-dipolar interactions $l_i(r_{i})$, $l_i^{\mathrm{off}}(r_{i})$. The parameters are related to magnetoelastic behavior depicted in the Figure~\ref{fig:bi_plots}. The pseudo-dipolar interaction is applied only to the Fe sublattice in the presented SL model. }
\begin{ruledtabular}
\begin{tabular}{c|ccc}
  & $\alpha$ (meV) & $\gamma$ & $\delta$ (\AA) \\

\hline
 &\multicolumn{3}{c}{SL model without  the $l_4$ interaction}\\  
$l_1$ &-5.087232 &	0.146938 & 2.722735 \\
$l_2$ & 7.146105 &	-0.372690 & 3.768487 \\
$l_3$ &0.186877	 & 1.000000	& 3.850528 \\
  \hline
 &\multicolumn{3}{c}{SL model with only the $l_4$ interaction}\\  
$l_4$ & 15.079223 &1.000000 &4.649170 \\
$l_2^{\mathrm{off}}$ & -6.795979& 	1.000000 &	3.768487   \\
\hline
&\multicolumn{3}{c}{SL model with all $l_{i}$ interactions}\\
$l_1$ & -8.634783	&0.497414	&2.722735 \\
$l_2$ & 15.088129&	-1.451815	&3.768487\\
$l_3$ & 0.186877 &	1.000000	& 3.850528\\
$l_4$ & 15.079223 &	1.000000	& 4.649170\\
$l_2^{\mathrm{off}}$ &	-0.006796979	 & 1.000000	 &3.768487 \\
\hline
&\multicolumn{3}{c}{SL model with rescaled $b_{21}$}\\
$l_1$ & -7.570518	&0.426760	&2.722735 \\
$l_2$ & 12.705522&	-1.269732	&3.768487 \\
$l_3$ &  0.186877	&1.000000	&3.850528 \\
$l_4$ & 10.555456	&1.000000	&4.649170 \\
$l_2^{\mathrm{off}}$ &		-4.757186	&1.000000	&3.768487 \\

\end{tabular}
\end{ruledtabular}
\end{table}

 \section{Calculation details}

Background ab-initio calculations, including structure relaxation and calculation of magnetocrystalline anisotropy energy (MAE),  elastic and magnetoelastic parameters  used for setting the numerical parameters of the spin lattice-model were done within Vienna ab-initio simulation package (VASP)~{\cite{Kresse_VASP_r96,Kresse_VASP_r99}} employing projector-augmented-wave (PAW) method with PAW pseudo-potentials.
Non-collinear spin calculations including spin-orbit coupling consider the generalized gradient approximation (GGA) of Perdew-Burke-Ernzerhof (PBE)~\cite{PBE_r96} with the plane wave cut-off of 450~eV for the MAE and elastic constants calculations, resp. 520~eV in case of the magnetoelastic parameters.  Non-spherical contributions to the gradient corrections were included. Calculations were performed on an automatic $k$-mesh  with $R_{k}$~=~70 by tetrahedron Brillouin zone integration, except for magnetoelastic parameters, while higher $R_{k}$~=~130 and  Methfessel-Paxton scheme with a smearing 0.01~eV was employed.  The same energy smearing was used for the structure relaxation. In general, energy convergence better than 10$^{-6}$~eV was considered. Regarding MAE and magnetoelastic parameters more accurate convergence better than 10$^{-9}$~eV  was required. The elastic and magnetoelastic parameters were estimated by a finite displacements method employing  
AELAS~\cite{AELAS_r17} and MEALAS~\cite{MAELAS_2_r22} packages, serving for the generation of distorted structures, including the spin direction for magnetoelastic constants, and data analysis.

Besides, the Relativistic Spin Polarized toolkit (RSPt) package was used to evaluate the isotropic magnetic exchange interaction parameters~\cite{Wills2010_lmto_rspt,Kvashnin_prb91_r15}
considering the fully relativistic LKAG-based method~\cite{Szilva_r23_RevModPhys}. Fully relativistic calculations employing  VASP relaxed structure parameters consider the xc-potential of PBE 1996~\cite{PBE_r96} and Perdew Wang 1992~\cite{PW92_prb45_r92}, 25$\times$25$\times$25 $k$-mesh, and energy convergence was better than 10$^{-10}$~Ry.

Spin-lattice model simulations were performed within the LAMMPS package~\cite{LAMMPS}, where the above-introduced tetragonal spin-lattice model was implemented in the LAMMPS spin package~\cite{Tranchida_jpc_18_spinpackage}. In the simulations, a FePt interatomic potential of the MEAM type~\cite{Kim_jmr_r06_FePtMEAM} was considered as well as a custom interatomic potential in the  reference-free modified embedded atom method (RF-MEAM)\cite{r15_Duff_CompPhysComm,r22_Slooter_JPCM} framework. The custom RF-MEAM potential was based on the ab-initio VASP results considering 102 distorted L1$_{0}$ FePt structures.

\begin{table}[b]
\caption{\label{Tab:MAE_param}%
SL model values of MCA constants $K_{i}$ depending on the type of included pseudo-dipolar interaction $l_{i}$. The MCA is applied only to the Fe sublattice in the presented SL model. }
\begin{ruledtabular}
\begin{tabular}{ccc}
   $\tilde{K}_{1}$ (meV/f.u.) & ${K}_{2}$ (meV/f.u.) & ${K}_{3}$ (meV/f.u.) \\
\hline
 \multicolumn{3}{c}{SL model without  the $l_4$ interaction}\\ 
 18.06475 & 0.00000 & 0.01525 \\
  \hline
 \multicolumn{3}{c}{SL model with only the $l_4$ interaction}\\  
 -2.75592 & 0.0000 & 0.01525 \\
\hline
\multicolumn{3}{c}{SL model with all $l_{i}$ interactions}\\
58.06637 & 0.00000 & 0.01525 \\
\hline
\multicolumn{3}{c}{SL model with rescaled $b_{21}$}\\

46.06588 &  0.00000 & 0.01525 \\

\end{tabular}
\end{ruledtabular}
\end{table}

\begin{table}[b]
\caption{\label{Tab:Jij_Bethe-Slater_param}%
Bethe-Slater curve parameters for Fe-Fe magnetic exchange interaction $J_i(r_{i})$. }
\begin{ruledtabular}
\begin{tabular}{c|ccc}
  & $\alpha$ (meV) & $\gamma$ & $\delta$ (\AA) \\

\hline
$J_1$ & 10.054688 & 0.650589 & 2.722733 \\
$J_2$ & 1.450424 & 0.035201 & 3.768488 \\
$J_3$ & 8.624737 & 0.262803 & 3.850526 \\
$J_4$ & -0.280298 & 6.276936 & 4.649171 \\
$J_5$ & -3.269634 & 0.005610 & 5.387770 \\
\end{tabular}
\end{ruledtabular}
\end{table}

\begin{table}[b]
\caption{\label{Tab:magel_prop}%
Magnetoelastic constants of L1$_{0}$ FePt were obtained either by linear interpolation of the energy versus strain data given either by ab-initio calculations or spin-lattice (SL) model simulation in LAMMPS (Fig.~\ref{fig:bi_plots}).   }
\begin{ruledtabular}
\begin{tabular}{c|cc|cc}
 &\multicolumn{2}{c|}{VASP} & \multicolumn{2}{c}{SL model}  \\
 & &  &nonscaled&rescaled \\

\hline
 & (MPa) & (meV/f.u.) &  \multicolumn{2}{c}{(meV/f.u.)} \\
\hline

b$_{21}$&  \phantom{-}83.62&    \phantom{-}14.579 & \phantom{-}20.825 & \phantom{-}14.825 \\
b$_{22}$ & -44.96   &    \phantom{0}-7.838 & \phantom{0}-7.837 & \phantom{0}-7.837 \\
b$_{3}$ & -12.62   &    \phantom{0}-2.200 & \phantom{0}-2.199 & \phantom{0}-2.199 \\
b$_{4}$ & -44.59   &    -15.547 & -16.105 & -16.103 \\
b$_{3}^{\prime}$ & \phantom{-}76.41   &   \phantom{-}26.644 & \phantom{-}26.642 & \phantom{-}26.642 \\

\end{tabular}
\end{ruledtabular}
\end{table}

\section{Spin-lattice model for  tetragonal(I) F\MakeLowercase{e}P\MakeLowercase{t} system.}

\begin{figure}
\includegraphics[width=0.85\columnwidth]{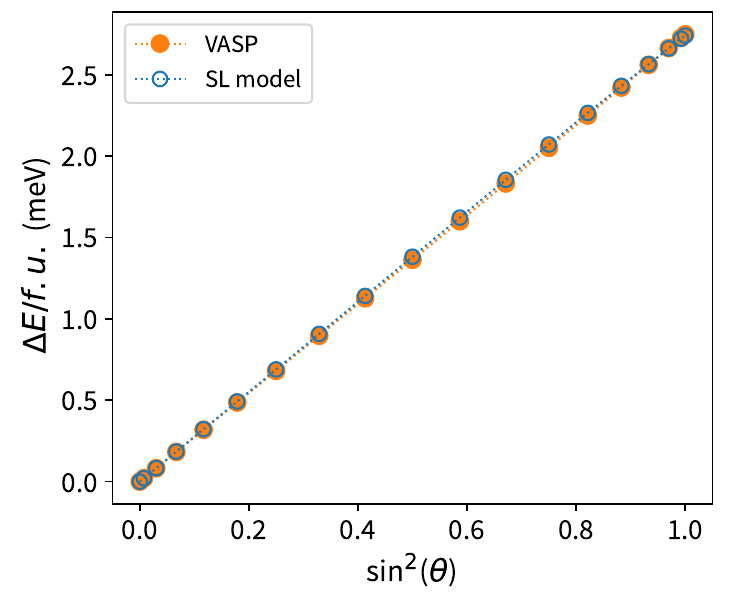}
\includegraphics[width=0.85\columnwidth]{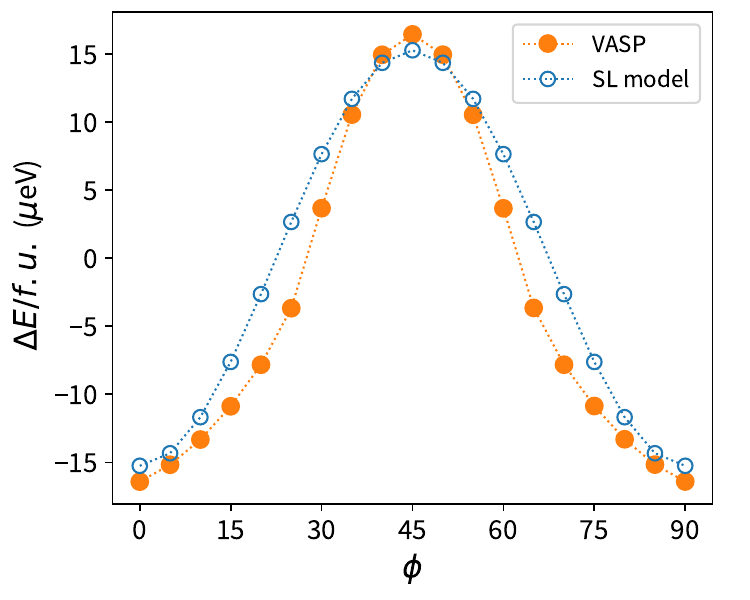}
\caption{\label{Fig:vasp_sl_mae} Magnetocrystalline anisotropy simulation for FePt system obtained by ab initio VASP calculations (filled points) and by the SL-model simulations (empty points). }
\end{figure}

To manifest the behavior of the above proposed SL model for  tetragonal systems, the SL model was used to reproduce the magnetoelastic behavior as calculated by ab-initio VASP calculation. Therefore, initially, prior the SL model simulations,  elastic and magnetoelastic properties were determined by ab-initio VASP calculations, using relaxed structure parameters in Table~\ref{Tab:elast_prop} , except the magnetoelastic calculations where the a finer $k$-mesh was used to achieve a linear behavior of magnetization direction dependent energy difference under small strain  ($a$=2.722~\AA, $c$=3.770~\AA). The obtained lattice parameters, elastic (Table~\ref{Tab:elast_prop}) and magnetoelastic constants correspond well to the literature~\cite{Ravindran_PRB_01,Lu_r10_prb_fept,Nieves_sss_r25_seconorder}. Besides, the ab-initio MCA constants were estimated (Fig.~\ref{Fig:vasp_sl_mae}), where $K_{1}$=15.805~MJ/m$^3$, $K_{2}\sim$0~MJ/m$^3$, and  $K_{3}$=0.087~MJ/m$^3$. 

\begin{figure}
\includegraphics[width=\columnwidth]{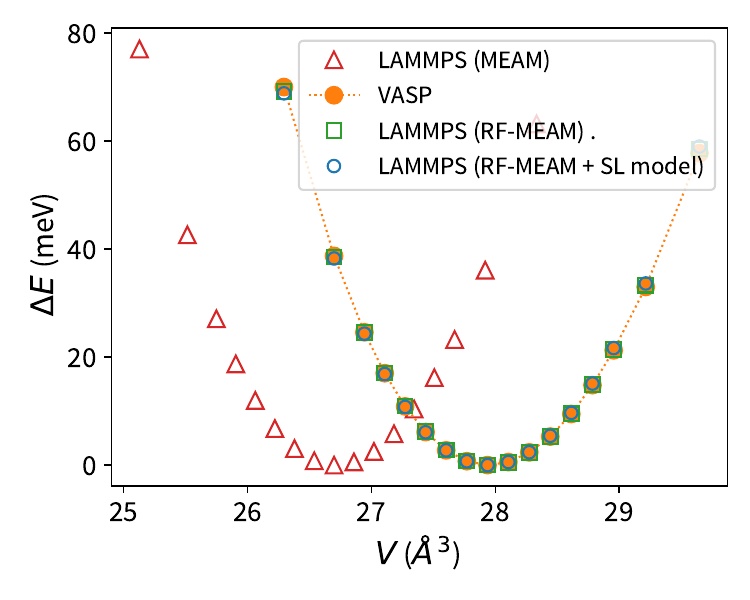}
\caption{\label{fig:eos} Volume dependence of the total energy difference in FePt calculations. (filled point) DFT result calculated in VASP. (empty triangle) LAMMPS simulation with an alloy potential. (empty square) LAMMPS simulation with DFT-based potential, (empty circle) LAMMPS simulation with DFT-based potential and  SL model included.    }
\end{figure}

The ab-initio calculations provide the following magnitudes of spin magnetic moments: $\mu_{S}^{Fe}$=2.92~$\mu_{B}$ and $\mu_{S}^{Pt}$=0.31~$\mu_{B}$ (compare to the value of Ref.~\cite{Lu_r10_prb_fept}), where $\mu_{B}$ denotes the Bohr magneton. Assuming the dipole-dipole interaction,
\begin{align}
    &\mathcal{H}_{dd}(\mathbf{r},\bm{\mu}) \propto \Big[ (\mathbf{e}_{ij}\cdot\bm{\mu}_{i})(\mathbf{e}_{ij}\cdot\bm{\mu}_{j}) - \frac{(\bm{\mu}_{i}\cdot\bm{\mu}_{j})}{3} \Big]\, , \nonumber
\end{align}
where the actual magnitudes of the magnetic moments occur ($\bm{\mu}_{i}$ denotes spin moment vector at the site $i$), unlike the pseudo-dipolar interaction (Eq.~\ref{Eq.pseudodip_ham}), the Fe-Pt resp. Pt-Pt dipolar interactions will be one order resp. two orders of magnitude smaller than the dipolar one for Fe-Fe interaction. Therefore, the SL model considers only the Fe sublattice as an approximation.  For simplicity, it is applied not only for the pseudo-dipolar interaction but also for the magnetic exchange and MCA interactions (Eq.~\ref{Eq.Ham_magnetic}). Otherwise, the pseudo-dipolar interactions of the proposed SL model can be applied separately to the Fe and Pt sublattices, and the same can be used for the MCA interactions.

\begin{figure}
\includegraphics[width=\columnwidth]{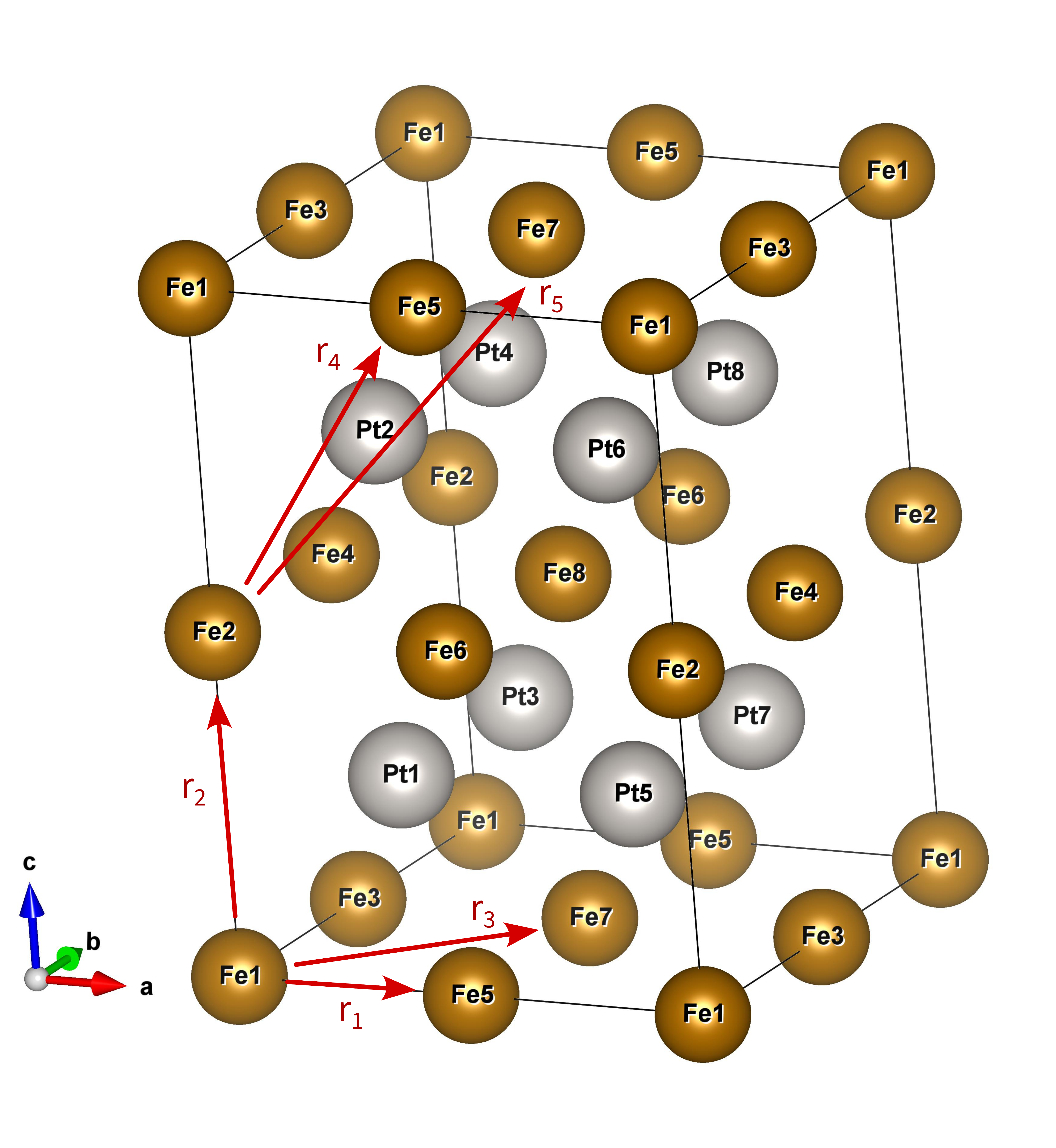}
\caption{FePt 2$\times$2$\times$2-supercell. Exchange interaction directions for the five nearest Fe-Fe neighbors depicted.  (plotted in VESTA 3~\cite{VESTA})}
\label{Fig.:suppercell222_FePt}
\end{figure}

\begin{figure}
\includegraphics[width=\columnwidth]{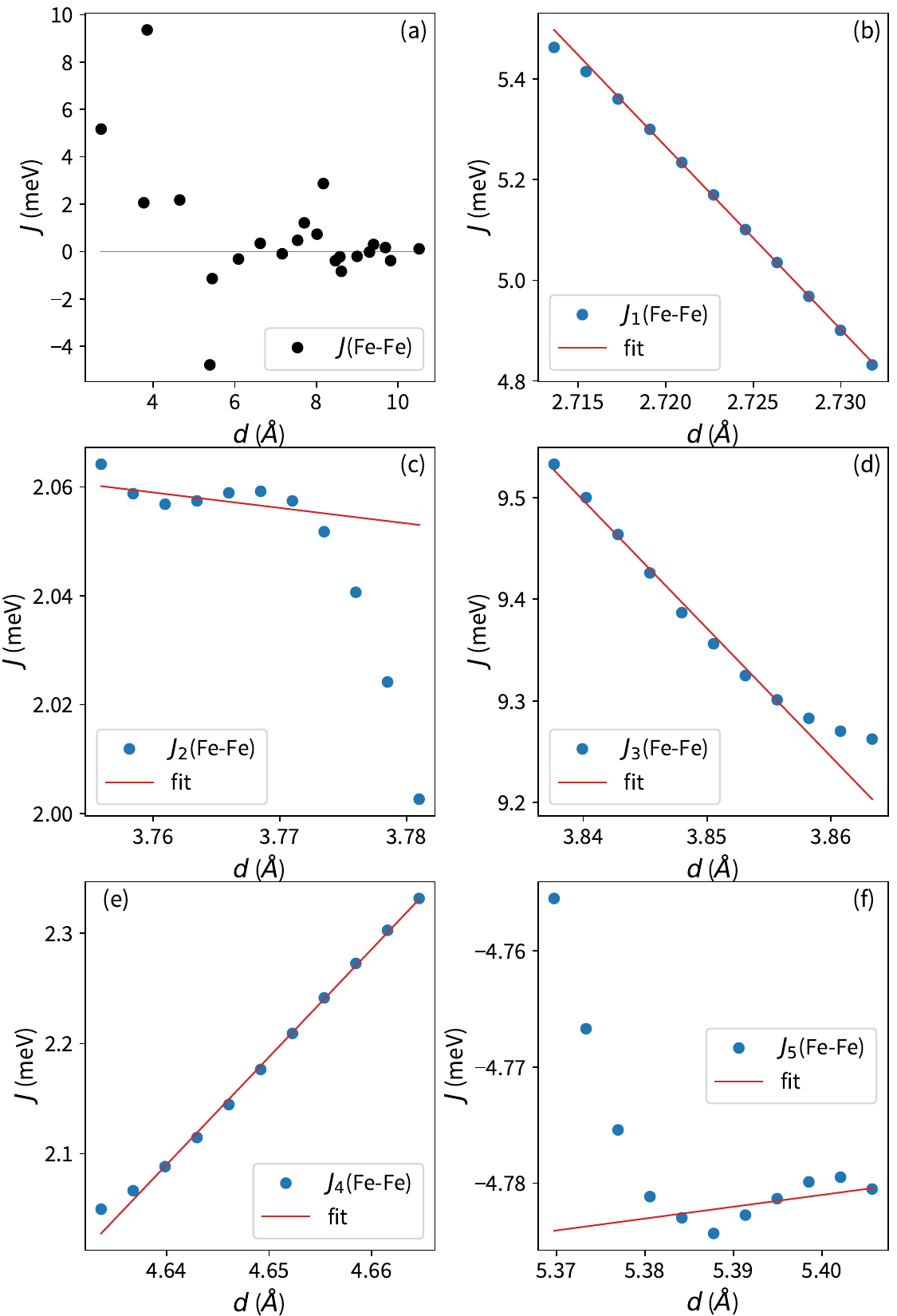}
\caption{\label{fig:jij} FePt magnetic exchange coupling parameters. (a) radial dependance of the exchange coupling $J(d)$. (b-f)  Exchange coupling $J_{i}(d)$ under volume change of the FePt cell. }
\end{figure}

Building the FePt SL model, initially, the elastic behavior given only by the interatomic potential was tested by neglecting the magnetic part $\mathcal{H}_{M}$ in the SL Hamiltonian (Eq.~\ref{Eq.SL_Hamiltonian}). Evaluating the elastic constants by the finite-displacements method  and the energy vs. volume curve and their comparison with the ab-initio results (Table~\ref{Tab:elast_prop}) reveals that the publicly available alloy-based FePt MEAM potential~\cite{Kim_jmr_r06_FePtMEAM} does not reproduce the values with high accuracy. Indeed, it exhibits a significant discrepancy in the elastic constants, particularly, for the $C_{12}$ constant that differs by more than 120\%, see Table~\ref{Tab:elast_prop}. Besides, the equilibrium volume is modified by 4\%. Therefore, in the SL model, a custom ab-initio data-based, interatomic potential of RF-MEAM~\cite{r15_Duff_CompPhysComm} type was developed. It reproduces well the ab initio energy-volume curve (Fig.~\ref{fig:eos}) as well as the elastic constants $C_{ij}$ (Table~\ref{Tab:elast_prop}).

\begin{figure*}
\includegraphics[width=\textwidth]{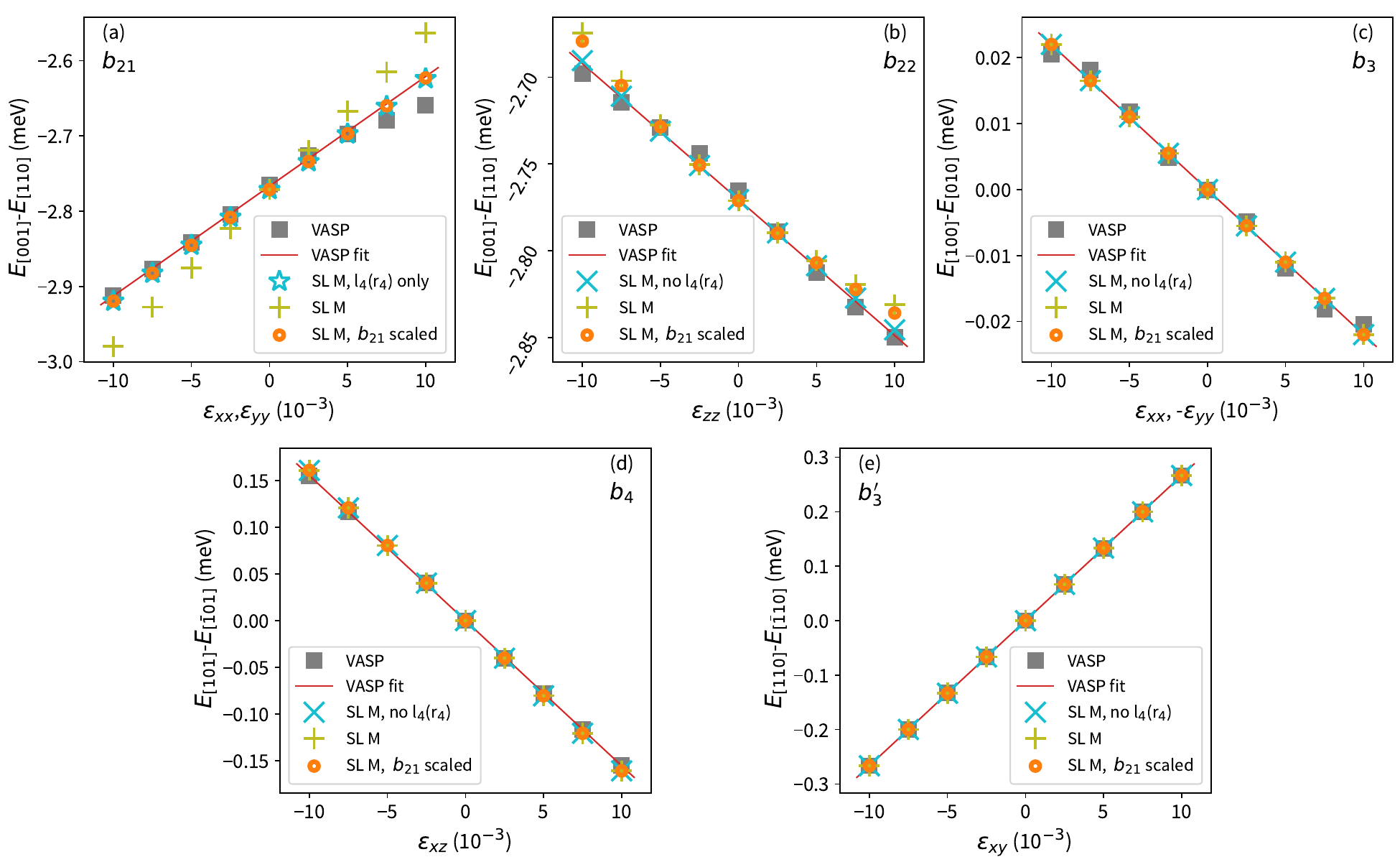}

\caption{\label{fig:bi_plots}  Energy versus strain data for estimation of the magnetoelastic  constants $b_i$. (square) ab-initio VASP results, (lines) linear fit of the VASP results, (star) spin-lattice model (SLM) with only $l_{4}(r_{4})$ interaction included , (crossed diagonal) SLM where $l_{4}(r_{4})$ interaction is excluded, (cross) SLM including all pseudo-dipolar interactions, (circle) SLM with rescaled $b_{21}$. The direction in the subscripts of the energy axes denotes the direction of the magnetization axis in FM-ordered FePt system.   }
\end{figure*}

To mimic the anisotropic magnetoelastic behavior described by the magnetoelastic energy (Eq.~\ref{Eq.E_mag-elast}), pseudo-dipolar interactions $l_{i}(r_{i})$ as derived above (Eq.~\ref{Eq.psdip_l1}-Eq.~\ref{Eq.psdip_dl4dr}) were introduced in the    SL-Hamiltonian (Eq.~\ref{Eq.SL_Hamiltonian}) based on the ab-initio magnetoelastic constants $b_{i}$. It also includes the $l_{2}^{off}$ interaction to off-set extra term in arising from $l_{4}$ interaction parametrization.   The  pseudo-dipolar interactions $l_{i}(r_{i})$ were described by Bethe-Slater curves (Eq.~\ref{Eq.Bethe-Salpeter_curve}) with parameters given in the Table~\ref{Tab:bi_Bethe-Slater_param}.
Depending on the included pseudo-dipolar interactions $l_{i}(r_{i})$, the MCA constants (Eq.~\ref{Eq.MAE_Ham_sn}) of the SL model have to be adapted (Table~\ref{Tab:MAE_param}), namely the leading $K_{1}$ (Eq.~\ref{Eq.MAE_K1_off}), to offset the contribution of the $l_{i}$ interactions to provide correct MAE (Fig.~\ref{Fig:vasp_sl_mae}). 
As mentioned, for simplicity, the magnetic exchange interaction $\mathcal{H}_{ex}$ (Eq.~\ref{Eq.SL_Hamiltonian}) was included only for the Fe sublattice (up to the the fifth nearest neighbor interaction $J_{i}(r_{i})$ were considered (Fig.~\ref{Fig.:suppercell222_FePt})), while the Pt ones were neglected. To approximate the radial dependence of the exchange coupling parameters $J_{i}$ (Eq.~\ref{Eq.SL_Hamiltonian}), a linear interpolation of the $J_{ij}$ volume dependence near the equilibrium volume was performed (Fig~\ref{fig:jij}). Parameterizing the  exchange coupling  $J_{ij}(r_{ij})$ by Bethe-Slater curves (Eq.~\ref{Eq.Bethe-Salpeter_curve}),  parameters $\alpha$, $\gamma$, $\delta$ in Table~\ref{Tab:Jij_Bethe-Slater_param} were used to describe the exchange interaction $\mathcal{H}_{ex}$.

Applying pseudo-dipolar interactions $l_{i}(r_{i})$ one by one, the SL-model simulate properly the ab-initio behavior when $l_{1}$-$l_{3}$ interactions (Fig.~\ref{fig:bi_plots}b-e, diagonal crosses) were used separately from the $l_{4}$ (Fig.~\ref{fig:bi_plots}a, stars). The energy versus strain data follow the linear interpolation of the VASP calculated data (Fig.~\ref{fig:bi_plots}, lines), Combining all the $l_{i}$ interactions together (Fig.~\ref{fig:bi_plots}, crosses), the SL-model start to overestimate the $b_{21}$ parameter as the slope of the SL model data  (Fig.~\ref{fig:bi_plots}a, crosses) was stepper than that one of the linear fit. Further a quadratic-like extra contribution occurs for $b_{22}$ data (Fig.~\ref{fig:bi_plots}b, crosses). It suggests an influence of terms beyond the performed approximation of the strain modification of pseudo-dipolar interaction (Eq.~\ref{Eq.PsDipolarDiff}). To restore the ab initio like magnetoelastic behavior, the VASP calculated $b_{21}$ magnetoelastic constant used for the parametrization of the pseudo-dipolar interactions  (Eq.~\ref{Eq.psdip_l1}-Eq.~\ref{Eq.psdip_dl4dr}) was arbitrarily scaled by a factor $f_{b21}$=0.7. Employing the SL-model with rescaled $l_{i}$ interaction, the proper magnetoelastic behavior was mostly restored  (Fig.~\ref{fig:bi_plots}, circles). Only a noticeable quadratic contribution was kept in the determination of the $b_{22}$ magnetoelastic constant  (Fig.~\ref{fig:bi_plots}b). The values of the magnetoelastic constants $b_{i}$ determined by linear interpolation of data from the final rescaled SL-model well correspond to the ab initio values (Table~\ref{Tab:magel_prop}), with the maximum difference of 3.6\% for the $b_{4}$ constant.

Further, the proposed SL model reproduces well the MCA behavior following the ab initio MAE data (Fig.~\ref{Fig:vasp_sl_mae}). Using the properly scaled $\tilde{K}_{1}$ constant (Table~\ref{Tab:MAE_param}), the correct magnitude of the MCA is given by the SL model, irrespective of the $l_{i}$ interactions included, as manifested by zero strain data in Fig.~\ref{fig:bi_plots}a,b, which provides the energy difference between the [001] and [110] magnetization directions.

\section{Conclusions}
In summary, a spin lattice model for tetragonal symmetry was introduced and tested for a representative system of FePt. By analyzing the behavior of the magnetic terms in the spin-lattice Hamiltonian, the parametrization of the magnetic interactions for a tetragonal system was proposed. Namely, a simple model of the pseudo-dipolar interactions reproducing the anisotropic magnetoelastic behavior prescribed by the magnetoelastic energy formula was derived. Including the magnetic exchange interaction and magneto-crystalline anisotropy in the spin lattice Hamiltonian, corrections eliminating unwanted interference were proposed. 

The derived tetragonal spin-lattice model was implemented in the LAMMPS package and tested for a L1$_{0}$ FePt -- a characteristic representative of tetragonal (I) systems. It was proven that the proposed spin-lattice model reproduces the behavior obtained by ab-initio calculations. In addition, we have developed a custom DFT-based data interaction potential for spin-lattice simulations that describes the energy-volume as well as the elastic constants ($C_{ij}$) with much higher accuracy than before.

\FloatBarrier
\section{Acknowledgemets}
{This work has been supported by GAČR project  24-11388I of the Grant Agency of Czech Republic and by the Ministry of Education, Youth
and Sports of the Czech Republic through the e-INFRA CZ (ID:90254)
}

\section*{Author contributions}

\textbf{JŠ}: Conceptualization, Methodology, Software, Data curation, Formal analysis, Investigation, Visualization, Writing – original draft; \textbf{DL}:  Formal analysis, Writing – review \& editing

\FloatBarrier

\bibliography{apssamp}

@PREAMBLE{
 "\providecommand{\noopsort}[1]{}" 
 # "\providecommand{\singleletter}[1]{#1}%" 
}

@article{Lu_r10_prb_fept,
  title = {First-principles study of magnetic properties of $\text{L}{1}_{0}$-ordered {MnPt} and FePt alloys},
  author = {Lu, Zhihong and Chepulskii, Roman V. and Butler, W. H.},
  journal = {Phys. Rev. B},
  volume = {81},
  issue = {9},
  pages = {094437},
  numpages = {8},
  year = {2010},
  month = {Mar},
  publisher = {American Physical Society},
  doi = {10.1103/PhysRevB.81.094437},
  url = {https://link.aps.org/doi/10.1103/PhysRevB.81.094437}
}

@article{r22_Slooter_JPCM,
doi = {10.1088/1361-648X/ac9d14},
url = {https://dx.doi.org/10.1088/1361-648X/ac9d14},
year = {2022},
month = {nov},
publisher = {IOP Publishing},
volume = {34},
number = {50},
pages = {505901},
author = {Slooter, Rutger J and Sluiter, Marcel H F and Kranendonk, Winfried G T and Bos, Cornelis},
title = {A reference-free {MEAM} potential for {$\alpha$}-{Fe} and {$\gamma$}-{Fe}},
journal = {Journal of Physics: Condensed Matter}
}

@article{r15_Duff_CompPhysComm,
title = {{MEAM}fit: A reference-free modified embedded atom method ({RF}-{MEAM}) energy and force-fitting code},
journal = {Computer Physics Communications},
volume = {196},
pages = {439-445},
year = {2015},
issn = {0010-4655},
doi = {https://doi.org/10.1016/j.cpc.2015.05.016},
url = {https://www.sciencedirect.com/science/article/pii/S0010465515001964},
author = {Andrew Ian Duff and M.W. Finnis and Philippe Maugis and Barend J. Thijsse and Marcel H.F. Sluiter}
}

@Article{Kim_jmr_r06_FePtMEAM,
author={Kim, Jaesong
and Koo, Yangmo
and Lee, Byeong-Joo},
title={Modified embedded-atom method interatomic potential for the {F}e--{P}t alloy system},
journal={Journal of Materials Research},
year={2006},
month={Jan},
day={01},
volume={21},
number={1},
pages={199-208},
issn={2044-5326},
doi={10.1557/jmr.2006.0008},
url={https://doi.org/10.1557/jmr.2006.0008}
}

@Article{LAMMPS,
  author = "A. P. Thompson and H. M. Aktulga and R. Berger and 
     D. S. Bolintineanu and W. M. Brown and P. S. Crozier and
     P. J. in 't Veld and A. Kohlmeyer and S. G. Moore and T. D. Nguyen and
     R. Shan and M. J. Stevens and J. Tranchida and C. Trott and S. J. Plimpton",
  title = "{LAMMPS} - a flexible simulation tool for
     particle-based materials modeling at the 
     atomic, meso, and continuum scales",
  journal = "Comp. Phys. Comm.",
  volume =  "271",
  pages =   "108171",
  year =    "2022",
  doi = "10.1016/j.cpc.2021.108171"
}

@article{Korniilenko_ResPhys_r25,
title = {Understanding change in the sound wave frequency in a ferromagnet under magnetic field influence ({S}imon effect) in the low-field regime},
journal = {Results in Physics},
volume = {73},
pages = {108264},
year = {2025},
issn = {2211-3797},
doi = {https://doi.org/10.1016/j.rinp.2025.108264},
url = {https://www.sciencedirect.com/science/article/pii/S2211379725001585},
author = {I. Korniienko and P. Nieves and D. Legut}
}

@article{Tranchida_jpc_18_spinpackage,
title = {Massively parallel symplectic algorithm for coupled magnetic spin dynamics and molecular dynamics},
journal = {Journal of Computational Physics},
volume = {372},
pages = {406-425},
year = {2018},
issn = {0021-9991},
doi = {https://doi.org/10.1016/j.jcp.2018.06.042},
url = {https://www.sciencedirect.com/science/article/pii/S0021999118304200},
author = {J. Tranchida and S.J. Plimpton and P. Thibaudeau and A.P. Thompson}
}

@article{Korniienko_PRR_r24,
  title = {Computational study of elastic waves generated by ultrafast demagnetization in fcc Ni},
  author = {Korniienko, I. and Nieves, P. and Fraile, A. and Iglesias, R. and Legut, D.},
  journal = {Phys. Rev. Res.},
  volume = {6},
  issue = {2},
  pages = {023311},
  numpages = {11},
  year = {2024},
  month = {Jun},
  publisher = {American Physical Society},
  doi = {10.1103/PhysRevResearch.6.023311},
  url = {https://link.aps.org/doi/10.1103/PhysRevResearch.6.023311}
}

@article{Nieves_prb_21_sl_model,
  title = {Spin-lattice model for cubic crystals},
  author = {Nieves, P. and Tranchida, J. and Arapan, S. and Legut, D.},
  journal = {Phys. Rev. B},
  volume = {103},
  issue = {9},
  pages = {094437},
  numpages = {16},
  year = {2021},
  month = {Mar},
  publisher = {American Physical Society},
  doi = {10.1103/PhysRevB.103.094437},
  url = {https://link.aps.org/doi/10.1103/PhysRevB.103.094437}
}

@article{Cooke_cpc_r22_slmodel,
title = {An implicit spin lattice dynamics integrator in LAMMPS},
journal = {Computer Physics Communications},
volume = {271},
pages = {108203},
year = {2022},
issn = {0010-4655},
doi = {https://doi.org/10.1016/j.cpc.2021.108203},
url = {https://www.sciencedirect.com/science/article/pii/S0010465521003155},
author = {Joseph R. Cooke and Jennifer R. Lukes}
}

@article{Stockem_prl_r18_asdaimd,
  title = {Anomalous Phonon Lifetime Shortening in Paramagnetic CrN Caused by Spin-Lattice Coupling: A Combined Spin and Ab Initio Molecular Dynamics Study},
  author = {Stockem, Irina and Bergman, Anders and Glensk, Albert and Hickel, Tilmann and K\"ormann, Fritz and Grabowski, Blazej and Neugebauer, J\"org and Alling, Bj\"orn},
  journal = {Phys. Rev. Lett.},
  volume = {121},
  issue = {12},
  pages = {125902},
  numpages = {6},
  year = {2018},
  month = {Sep},
  publisher = {American Physical Society},
  doi = {10.1103/PhysRevLett.121.125902},
  url = {https://link.aps.org/doi/10.1103/PhysRevLett.121.125902}
}

@article{PETRIDIS2006131,
title = {A new magnetoelastic device for sensing applications},
journal = {Sensors and Actuators A: Physical},
volume = {129},
number = {1},
pages = {131-137},
year = {2006},
note = {EMSA 2004},
issn = {0924-4247},
doi = {https://doi.org/10.1016/j.sna.2005.11.033},
url = {https://www.sciencedirect.com/science/article/pii/S0924424705006746},
author = {C. Petridis and P. Dimitropoulos and E. Hristoforou}
}

@article{Bestok_micropump,
    author = {Beskok , A.  and Srinivasa , A. R. },
    title = {Simulation and Analysis of a Magnetoelastically Driven Micro-Pump},
    journal = {Journal of Fluids Engineering},
    volume = {123},
    number = {2},
    pages = {435-438},
    year = {2000},
    month = {10},
    issn = {0098-2202},
    doi = {10.1115/1.1363700},
    url = {https://doi.org/10.1115/1.1363700},
    eprint = {https://asmedigitalcollection.asme.org/fluidsengineering/article-pdf/123/2/435/5590470/435\_1.pdf},
}

@ARTICLE{garshelis_r99,
  author={Garshelis, I.J. and Jones, C.A.},
  journal={IEEE Transactions on Magnetics}, 
  title={Miniaturized magnetoelastic torque transducers}, 
  year={1999},
  volume={35},
  number={5},
  pages={3649-3651},
  doi={10.1109/20.800619}}

@ARTICLE{nishibe_r98_torquq,
  author={Nishibe, Y. and Nonomura, Y. and Tsukada, K. and Takeuchi, M. and Miyashita, M. and Ito, K.},
  journal={IEEE Transactions on Instrumentation and Measurement}, 
  title={Determination of engine misfiring using magnetoelastic torque sensor}, 
  year={1998},
  volume={47},
  number={3},
  pages={760-765},
  doi={10.1109/19.744343}}

@Inbook{MagnetoelasticWaves_book,
author="Baghdasaryan, Gevorg
and Danoyan, Zaven",
bookTitle="Magnetoelastic Waves",
year="2018",
publisher="Springer Singapore",
address="Singapore",
isbn="978-981-10-6762-4",
doi="10.1007/978-981-10-6762-4_1",
url="https://doi.org/10.1007/978-981-10-6762-4_1"
}

@article{VESTA,
author = "Momma, Koichi and Izumi, Fujio",
title = "{{\it VESTA3} for three-dimensional visualization of crystal, volumetric and morphology data}",
journal = "Journal of Applied Crystallography",
year = "2011",
volume = "44",
number = "6",
pages = "1272--1276",
month = "Dec",
doi = {10.1107/S0021889811038970},
url = {https://doi.org/10.1107/S0021889811038970},
}

@article{PW92_prb45_r92,
  title = {Accurate and simple analytic representation of the electron-gas correlation energy},
  author = {Perdew, John P. and Wang, Yue},
  journal = {Phys. Rev. B},
  volume = {45},
  issue = {23},
  pages = {13244--13249},
  numpages = {0},
  year = {1992},
  month = {Jun},
  publisher = {American Physical Society},
  doi = {10.1103/PhysRevB.45.13244},
  url = {https://link.aps.org/doi/10.1103/PhysRevB.45.13244}
}

@article{Kvashnin_prb91_r15,
  title = {Exchange parameters of strongly correlated materials: Extraction from spin-polarized density functional theory plus dynamical mean-field theory},
  author = {Kvashnin, Y. O. and Gr\aa{}n\"as, O. and Di Marco, I. and Katsnelson, M. I. and Lichtenstein, A. I. and Eriksson, O.},
  journal = {Phys. Rev. B},
  volume = {91},
  issue = {12},
  pages = {125133},
  numpages = {10},
  year = {2015},
  month = {Mar},
  publisher = {American Physical Society},
  doi = {10.1103/PhysRevB.91.125133},
  url = {https://link.aps.org/doi/10.1103/PhysRevB.91.125133}
}

@Inbook{Wills2010_lmto_rspt,
author="Wills, John M.
and Alouani, Mebarek
and Andersson, Per
and Delin, Anna
and Eriksson, Olle
and Grechnyev, Oleksiy",
title="The Full-Potential Electronic Structure Problem and {RSP}t",
bookTitle="Full-Potential Electronic Structure Method:  Energy and Force Calculations with Density Functional and Dynamical Mean Field Theory",
year="2010",
publisher="Springer Berlin Heidelberg",
address="Berlin, Heidelberg",
pages="47--73",
isbn="978-3-642-15144-6",
doi="10.1007/978-3-642-15144-6_6",
url="https://doi.org/10.1007/978-3-642-15144-6_6"
}

@article{maelas3_nieves_r23,
title = {Automated calculations of exchange magnetostriction},
journal = {Computational Materials Science},
volume = {224},
pages = {112158},
year = {2023},
issn = {0927-0256},
doi = {https://doi.org/10.1016/j.commatsci.2023.112158},
url = {https://www.sciencedirect.com/science/article/pii/S0927025623001520},
author = {P. Nieves and S. Arapan and S.H. Zhang and A.P. Kadzielawa and R.F. Zhang and D. Legut}
}

@article{Nieves_sss_r25_seconorder,
title = {Second-order anisotropy due to magnetostriction for {L}10-{F}e{P}t},
journal = {Solid State Sciences},
volume = {160},
pages = {107782},
year = {2025},
issn = {1293-2558},
doi = {https://doi.org/10.1016/j.solidstatesciences.2024.107782},
url = {https://www.sciencedirect.com/science/article/pii/S1293255824003479},
author = {D. Legut and P. Nieves}
}

@article{Ravindran_prb_r01,
  title = {Large magnetocrystalline anisotropy in bilayer transition metal phases from first-principles full-potential calculations},
  author = {Ravindran, P. and Kjekshus, A. and Fjellv\aa{}g, H. and James, P. and Nordstr\"om, L. and Johansson, B. and Eriksson, O.},
  journal = {Phys. Rev. B},
  volume = {63},
  issue = {14},
  pages = {144409},
  numpages = {18},
  year = {2001},
  month = {Mar},
  publisher = {American Physical Society},
  doi = {10.1103/PhysRevB.63.144409},
  url = {https://link.aps.org/doi/10.1103/PhysRevB.63.144409}
}

@Article{Spetzler_scirep_r21,
author={Spetzler, B.
and Bald, C.
and Durdaut, P.
and Reermann, J.
and Kirchhof, C.
and Teplyuk, A.
and Meyners, D.
and Quandt, E.
and H{\"o}ft, M.
and Schmidt, G.
and Faupel, F.},
title={Exchange biased delta-E effect enables the detection of low frequency pT magnetic fields with simultaneous localization},
journal={Scientific Reports},
year={2021},
month={Mar},
day={05},
volume={11},
number={1},
pages={5269},
issn={2045-2322},
doi={10.1038/s41598-021-84415-2},
url={https://doi.org/10.1038/s41598-021-84415-2}
}

@article{Calkins_jimss_r07,
author = {Frederick T. Calkins and Alison B. Flatau and Marcelo J. Dapino},
title ={Overview of Magnetostrictive Sensor Technology},

journal = {Journal of Intelligent Material Systems and Structures},
volume = {18},
number = {10},
pages = {1057-1066},
year = {2007},
doi = {10.1177/1045389X06072358},
URL = { 
        https://doi.org/10.1177/1045389X06072358
},
eprint = { 
        https://doi.org/10.1177/1045389X06072358
}
}

@article{Bienkovsky_SensAct_r04,
title = {The possibility of utilizing the high permeability magnetic materials in construction of magnetoelastic stress and force sensors},
journal = {Sensors and Actuators A: Physical},
volume = {113},
number = {3},
pages = {270-276},
year = {2004},
note = {New materials and Technologies in Sensor Applications, Proceedings of the European Materials Research Society 2003 - Symposium N},
issn = {0924-4247},
doi = {https://doi.org/10.1016/j.sna.2004.01.010},
url = {https://www.sciencedirect.com/science/article/pii/S0924424704000172},
author = {Adam Bieńkowski and Roman Szewczyk}
}

@article{Ravindran_PRB_01,
  title = {Large magnetocrystalline anisotropy in bilayer transition metal phases from first-principles full-potential calculations},
  author = {Ravindran, P. and Kjekshus, A. and Fjellv\aa{}g, H. and James, P. and Nordstr\"om, L. and Johansson, B. and Eriksson, O.},
  journal = {Phys. Rev. B},
  volume = {63},
  issue = {14},
  pages = {144409},
  numpages = {18},
  year = {2001},
  month = {Mar},
  publisher = {American Physical Society},
  doi = {10.1103/PhysRevB.63.144409},
  url = {https://link.aps.org/doi/10.1103/PhysRevB.63.144409}
}

@article{Szilva_r23_RevModPhys,
  title = {Quantitative theory of magnetic interactions in solids},
  author = {Szilva, Attila and Kvashnin, Yaroslav and Stepanov, Evgeny A. and Nordstr\"om, Lars and Eriksson, Olle and Lichtenstein, Alexander I. and Katsnelson, Mikhail I.},
  journal = {Rev. Mod. Phys.},
  volume = {95},
  issue = {3},
  pages = {035004},
  numpages = {71},
  year = {2023},
  month = {Sep},
  publisher = {American Physical Society},
  doi = {10.1103/RevModPhys.95.035004},
  url = {https://link.aps.org/doi/10.1103/RevModPhys.95.035004}
}

@book{chikazumi2009physics,
  title={Physics of Ferromagnetism},
  author={Chikazumi, S.},
  isbn={9780191569852},
  series={International Series of Monographs on Physics},
  url={https://books.google.cz/books?id=AZVfuxXF2GsC},
  year={2009},
  publisher={OUP Oxford}
}

@article{Kresse_VASP_r96,
  title = {Efficient iterative schemes for ab initio total-energy calculations using a plane-wave basis set},
  author = {Kresse, G. and Furthm\"uller, J.},
  journal = {Phys. Rev. B},
  volume = {54},
  issue = {16},
  pages = {11169--11186},
  numpages = {0},
  year = {1996},
  month = {Oct},
  publisher = {American Physical Society},
  doi = {10.1103/PhysRevB.54.11169},
  url = {https://link.aps.org/doi/10.1103/PhysRevB.54.11169}
}

@article{Kresse_VASP_r99,
  title = {From ultrasoft pseudopotentials to the projector augmented-wave method},
  author = {Kresse, G. and Joubert, D.},
  journal = {Phys. Rev. B},
  volume = {59},
  issue = {3},
  pages = {1758--1775},
  numpages = {0},
  year = {1999},
  month = {Jan},
  publisher = {American Physical Society},
  doi = {10.1103/PhysRevB.59.1758},
  url = {https://link.aps.org/doi/10.1103/PhysRevB.59.1758}
}

@article{PBE_r96,
  title = {Generalized Gradient Approximation Made Simple},
  author = {Perdew, John P. and Burke, Kieron and Ernzerhof, Matthias},
  journal = {Phys. Rev. Lett.},
  volume = {77},
  issue = {18},
  pages = {3865--3868},
  numpages = {0},
  year = {1996},
  month = {Oct},
  publisher = {American Physical Society},
  doi = {10.1103/PhysRevLett.77.3865},
  url = {https://link.aps.org/doi/10.1103/PhysRevLett.77.3865}
}

@article{AELAS_r17,
title = {AELAS: Automatic ELAStic property derivations via high-throughput first-principles computation},
journal = {Computer Physics Communications},
volume = {220},
pages = {403-416},
year = {2017},
issn = {0010-4655},
doi = {https://doi.org/10.1016/j.cpc.2017.07.020},
url = {https://www.sciencedirect.com/science/article/pii/S0010465517302400},
author = {S.H. Zhang and R.F. Zhang}
}

@article{MAELAS_1_r21,
title = {MAELAS: MAgneto-ELAStic properties calculation via computational hig--throughput approach},
journal = {Computer Physics Communications},
volume = {264},
pages = {107964},
year = {2021},
issn = {0010-4655},
doi = {https://doi.org/10.1016/j.cpc.2021.107964},
url = {https://www.sciencedirect.com/science/article/pii/S0010465521000801},
author = {P. Nieves and S. Arapan and S.H. Zhang and A.P. Kadzielawa and R.F. Zhang and D. Legut}
}

@article{MAELAS_2_r22,
title = {MAELAS 2.0: A new version of a computer program for the calculation of magneto-elastic properties},
journal = {Computer Physics Communications},
volume = {271},
pages = {108197},
year = {2022},
issn = {0010-4655},
doi = {https://doi.org/10.1016/j.cpc.2021.108197},
url = {https://www.sciencedirect.com/science/article/pii/S001046552100309X},
author = {P. Nieves and S. Arapan and S.H. Zhang and A.P. Kadzielawa and R.F. Zhang and D. Legut}
}

\end{document}